\documentclass[9pt,a4paper,twocolumn]{article}
\usepackage{iftex}
\usepackage[left=2cm,right=2cm,top=2.5cm,bottom=2.5cm]{geometry}
\usepackage{amsmath,amssymb,bm}
\usepackage{graphicx}
\usepackage{float}
\usepackage{dblfloatfix}
\usepackage{booktabs}
\usepackage{setspace}
\usepackage{titlesec}
\usepackage{caption}
\usepackage[numbers,sort&compress]{natbib}
\usepackage[hidelinks]{hyperref}
\ifPDFTeX
  \usepackage[T1]{fontenc}
  \usepackage[utf8]{inputenc}
  \usepackage{newtxtext,newtxmath}
  \usepackage[scaled=0.92]{helvet}
\else
  \usepackage{fontspec}
  \IfFontExistsTF{Times New Roman}{\setmainfont{Times New Roman}}{\setmainfont{TeX Gyre Termes}}
  \IfFontExistsTF{Arial}{\setsansfont{Arial}}{\setsansfont{TeX Gyre Heros}}
  \usepackage{newtxmath}
\fi

\DeclareCaptionFont{ninept}{\fontsize{9pt}{11pt}\selectfont}
\titleformat{\section}{\bfseries\fontsize{12pt}{15pt}\selectfont}{\thesection}{0.75em}{}
\titleformat{name=\section,numberless}{\bfseries\fontsize{12pt}{15pt}\selectfont}{}{0pt}{}
\titleformat{\subsection}{\bfseries\fontsize{11pt}{13.5pt}\selectfont}{\thesubsection}{0.75em}{}
\titleformat{name=\subsection,numberless}{\bfseries\fontsize{11pt}{13.5pt}\selectfont}{}{0pt}{}
\titlespacing*{\section}{0pt}{6pt}{6pt}
\titlespacing*{\subsection}{0pt}{6pt}{6pt}
\title{\textbf{Phase-slip residual-order spin state in FeSe}}

\author{Zhixin Liu$^{1}$,
Jiyu Fan$^{1,*}$,
Lei Zhang$^{3}$,
Chunlan Ma$^{4}$,
Daling Shi$^{1}$,
Zhongqin Yang$^{2,\dagger}$,
Yan Zhu$^{1,\ddagger}$
}

\date{}

\begin{document}

\twocolumn[
\begin{@twocolumnfalse}
\maketitle
\begin{center}
\small{$^{1}$College of Science, Nanjing University of Aeronautics and Astronautics, Nanjing 210016, China}\\
\small{$^{2}$Department of Physics, Fudan University, 200433 Shanghai, China}\\
\small{$^{3}$High Magnetic Field Laboratory, Chinese Academy of Sciences, Hefei 230031, China}\\
\small{$^{4}$Advanced Technology Research Institute of Taihu Photon Center, School of Physical Science and Technology, Suzhou University of Science and Technology, Suzhou, 215009, China}\\
\end{center}

\begin{abstract}
In unconventional superconductors, the microscopic form of magnetic correlations is crucial for identifying the origin of spin fluctuations and the associated pairing interaction. FeSe superconducts without chemical doping and shows no static long-range magnetic order, yet inelastic neutron scattering reveals strong stripe response, finite linewidths, and reproducible Néel-side spectral weight. Here we propose a phase-slip residual-order spin state (ROSS). Stripe, N\'eel, pair-checkerboard, and staggered trimer antiferromagnetic states can be unified as symmetric phase-slip derivatives of a stripe background; more general asymmetric phase slips form lower-energy configurations and reconstruct the inelastic neutron scattering spin spectrum $S(\mathbf q)$ within a finite coherence length. The ROSS therefore reconciles the absence of static magnetic order with strong spin excitations, provides a new microscopic picture for the origin of spin fluctuations in FeSe. The establishes a magnetic basis for understanding pairing in unconventional superconductors systems with similar magnetic fingerprints.
\end{abstract}
\vspace{1em}
\end{@twocolumnfalse}
]
\begingroup
\renewcommand{\thefootnote}{\fnsymbol{footnote}}
\footnotetext[1]{Contact author: jiyufan@nuaa.edu.cn}
\footnotetext[2]{Contact author: zyang@fudan.edu.cn}
\footnotetext[3]{Contact author: yzhu@nuaa.edu.cn}
\endgroup

{In the iron-based superconductors, FeSe occupies an unusual limit: it is structurally simple and superconducts without chemical doping \cite{ref1,ref2}, yet nuclear magnetic resonance (NMR) measurements find no static long-range magnetic order at ambient pressure \cite{ref3,ref4}, consistent with low-pressure zero-field $\mu$SR results \cite{ref5}.} This absence of order is not an absence of magnetism. Inelastic neutron scattering (INS) experiments reveal a highly organized spin spectrum: the response is dominated by stripe-channel weight, but it also carries weaker and reproducible spectral weight near the N\'eel wave vector \cite{ref6,ref7,ref8,ref9}. {The question is therefore not whether FeSe is magnetic, but what spin state produces structured momentum-space correlations without static long-range order in real space.}

{The weak N\'eel-channel weight is nontrivial not because it is large, but because it constrains the microscopic magnetic state. In spin-fluctuation-based descriptions, the assumed magnetic correlation pattern and the associated normal-state spin correlations provide the input from which the pairing interaction is constructed \cite{ref10,ref11,ref12,ref13}. It is also important for itinerant theory: because the spin susceptibility is derived from tight-binding band models, the relative spectral weight in the stripe and N\'eel channels directly tests the orbital content, Fermi-surface geometry, and quasiparticle coherence of the assumed electronic structure \cite{ref20,ref21}. The central problem is therefore to explain why a real N\'eel-like response coexists with stripe-dominated magnetism and no static long-range order.}

{Previous density-functional theory (DFT) calculations in periodic cells stabilize ordered magnetic states in FeSe \cite{ref14,ref15,ref16,ref17,ref18,ref19}, in apparent conflict with the absence of static long-range magnetic order at ambient pressure \cite{ref3,ref4,ref5}. Heisenberg-model-based descriptions reproduce selected aspects of the neutron spectrum, but they do not identify the real-space origin of the reproducible N\'eel-channel response \cite{ref22}. These results point beyond a single periodic magnetic order, toward a finite-coherence stripe state.}

\begin{figure}[t]
    \centering
    \includegraphics[width=\linewidth]{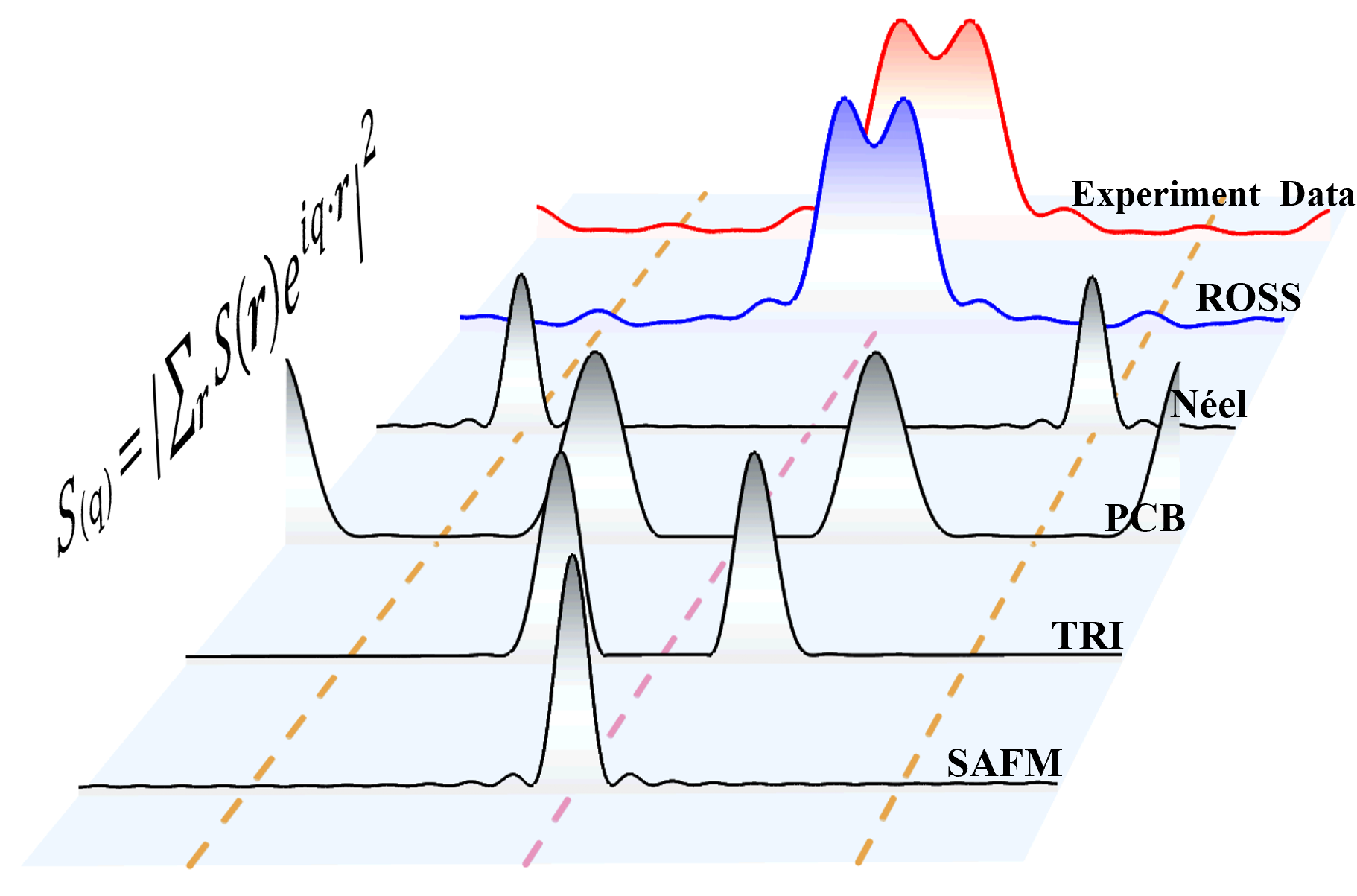}
    \caption{\textbf{Static spin structure factor $S(\mathbf q)$.}
Comparison of the ROSS signal with representative magnetic structures and the experimental $S(\mathbf q)$.}
    \label{fig:fig1}
\end{figure}

\begin{figure}[!b]
    \centering
    \includegraphics[width=\linewidth]{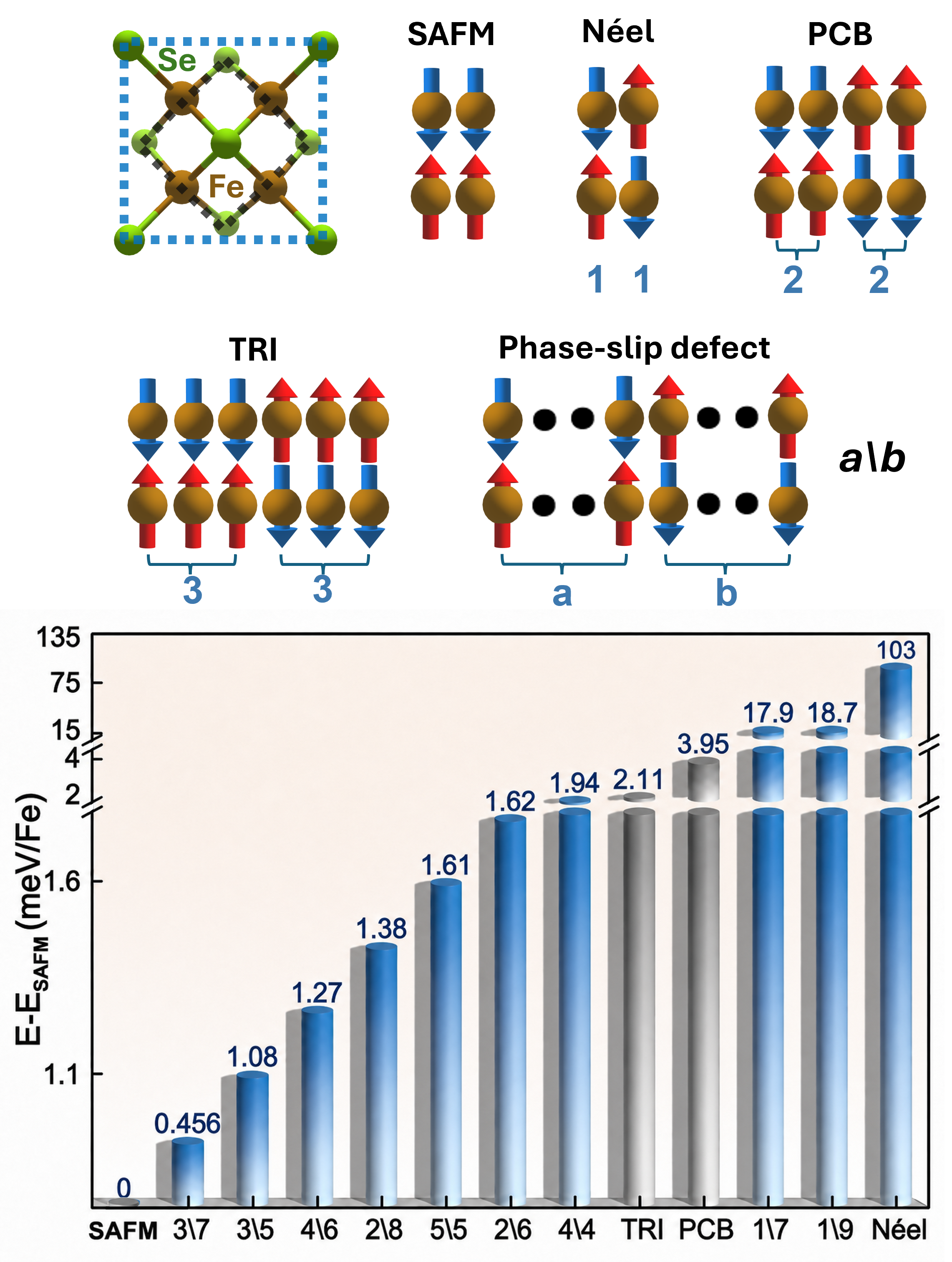}
    \caption{\textbf{Phase-slip magnetic manifold.}
(a) FeSe lattice, standard collinear states, and the generalized $a\backslash b$ phase-slip texture. (b) Relative energies $\Delta E=E-E_{\mathrm{SAFM}}$ (meV/Fe), showing that asymmetric phase-slip configurations form a low-energy manifold, with several members below the conventional TRI and PCB states, whereas N\'eel order remains high in energy on a broken y-axis.}
    \label{fig:fig2}
\end{figure}

To resolve this contradiction, we identify the magnetic state of FeSe as a phase-slip residual-order spin state (ROSS) [see Fig.~\ref{fig:fig1}].
{The ROSS denotes dynamic residual order: mobile $\pi$ phase slips reverse the stripe phase across slip walls, preserve stripe correlations within finite segments, and generate short-range N\'eel-like anticorrelations near the wall [see Fig.~\ref{fig:fig2}(a)].}
The resulting state is therefore distinct from a quantum spin liquid, a spin glass, and conventional disorder.
In this sense, the N\'eel-channel response is not the fingerprint of separate N\'eel order, but a local consequence of stripe phase slips in a ROSS background.

The static structure factor $S(\mathbf q)$ makes this distinction explicit [see Fig.~\ref{fig:fig1}]. Single periodic magnetic configurations,including stripe (SAFM), N\'eel, staggered-trimer (TRI)~\cite{ref14}, and pair-checkerboard (PCB)~\cite{ref15} antiferromagnetic orders, fail as complete descriptions of the observed spectrum. 
SAFM captures the dominant stripe channel but misses the residual N\'eel-side response; a genuine N\'eel state is incompatible with the absence of static order and with the dominant experimental line shape; and TRI and PCB carry characteristic peak-position and spectral mismatches. 
ROSS instead produces the required combination--a principal stripe response, finite linewidth, peak splitting, and a weak N\'eel-side component--as a single finite-coherence phase-slip signal.
The mapping [see Fig.~\ref{fig:fig2}(a)] then converts the usual list of collinear states into a single phase-slip sequence. SAFM is the parent stripe background, whereas N\'eel, PCB, and TRI correspond to different phase-slip densities [see Fig.~\ref{fig:fig2}(a)]. 
Thus the problem is no longer to choose one ordered state from a catalog of candidates, but to determine the low-energy landscape of stripe-derived phase slips.

\begin{figure*}[!t]
	\centering
	\includegraphics[width=0.80\textwidth]{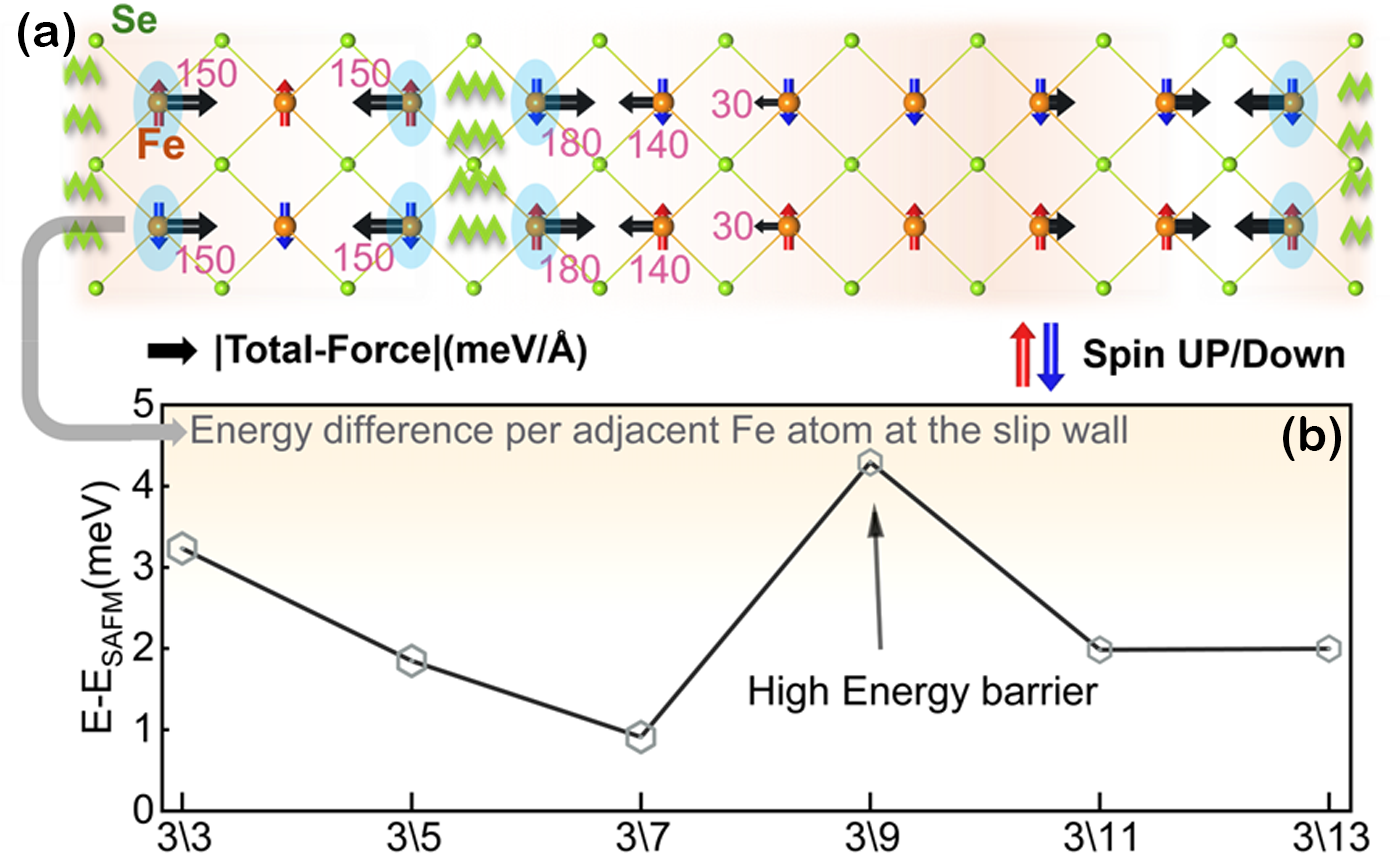}
	\caption{\textbf{Magneto-elastic response and slip-wall energetics.}(a) Force field induced by a $3\backslash 7$ slip wall on the relaxed SAFM lattice.(b) Effective slip-wall energy along the $3\backslash b$ phase-slip series, in which a three-site slip wall is embedded in stripe segments of length $b$. The energy is minimized near $3\backslash 7$ $N\approx 10$ and rises to a barrier near $3\backslash 9$.}
	\label{fig:fig3}
\end{figure*}

This viewpoint is further suggested by the TRI configuration [see Fig.~\ref{fig:fig2}(a)].
Although TRI is usually treated as a single periodic magnetic order, its real-space texture already contains inequivalent spin environments: moments near the slip-wall edge and moments inside the stripe segment are not equivalent.
This immediately implies that phase-slip textures should not be restricted to symmetric slip patterns.
Moreover, symmetric phase-slip configurations alone cannot reproduce the experimental $S(\mathbf q)$ line shapes, especially the simultaneous presence of a split or broadened stripe response and a weak N\'eel-side component.
{We therefore introduce asymmetric phase-slip textures.}

In the generalized $a\backslash b$ texture [see Fig.~\ref{fig:fig2}(a)], $a$ and $b$ denote the lengths of the two stripe segments separated by a slip wall, and the characteristic spatial scale is defined as $N=a+b$.
{This notation unifies SAFM as the zero-slip limit, N\'eel \((1\backslash1)\), PCB \((2\backslash2)\), TRI \((3\backslash3)\), and more general asymmetric textures [see Fig.~\ref{fig:fig2}(a)] into a single phase-slip family.}
The parameter $N$ controls the average spacing between neighboring slip walls, while the imbalance between $a$ and $b$ characterizes the asymmetry of the local phase-slip density.
By varying these two quantities, one can systematically examine how phase-slip defects reshape the magnetic response while preserving local stripe correlations.

{The relative energies in Fig.~\ref{fig:fig2}(b) were obtained using VASP with the mixed PBE--\(r^2\)SCAN exchange-correlation setup~\cite{ref23,ref24} described in Fig.S1.} This energy hierarchy provides the first central result. Several asymmetric $a\backslash b$ phase-slip textures such as \(3\backslash7\) and \(3\backslash5\) , not only remain close to SAFM, but lie below the conventional TRI and PCB states.
This establishes the phase-slip manifold as an intrinsic low-energy magnetic sector of FeSe, rather than a set of artificial interpolations between known collinear orders.
The sensitivity of this competing hierarchy to exchange-correlation mixing and DFT+$U$ is summarized in the Supplemental Material [see Fig.~S2].
In contrast, the N\'eel state is much higher in energy. Thus, the observed N\'eel-side response belongs to the slip-wall sector rather than to a competing long-range N\'eel phase; it can arise from local N\'eel-like fragments generated by phase slips in a stripe background.

To understand why the phase-slip configurations carry such a small energetic cost, we performed force calculations for a representative \(3\backslash7\) phase-slip configuration on the fully relaxed SAFM lattice background. The lattice response induced by the slip wall is distinctly nonlocal: the forces propagate through the elastic lattice network and remain appreciable up to the third-nearest-neighbor Fe atoms [see Fig.~\ref{fig:fig3}(a)]. This nonlocal response is the microscopic mechanism that makes ROSS viable in FeSe. Magnetoelastic coupling \cite{ref26,ref27} spreads the slip-wall formation energy over an extended region and strongly reduces the effective local penalty.
\begin{figure*}[t]
    \centering
    \includegraphics[width=0.95\textwidth]{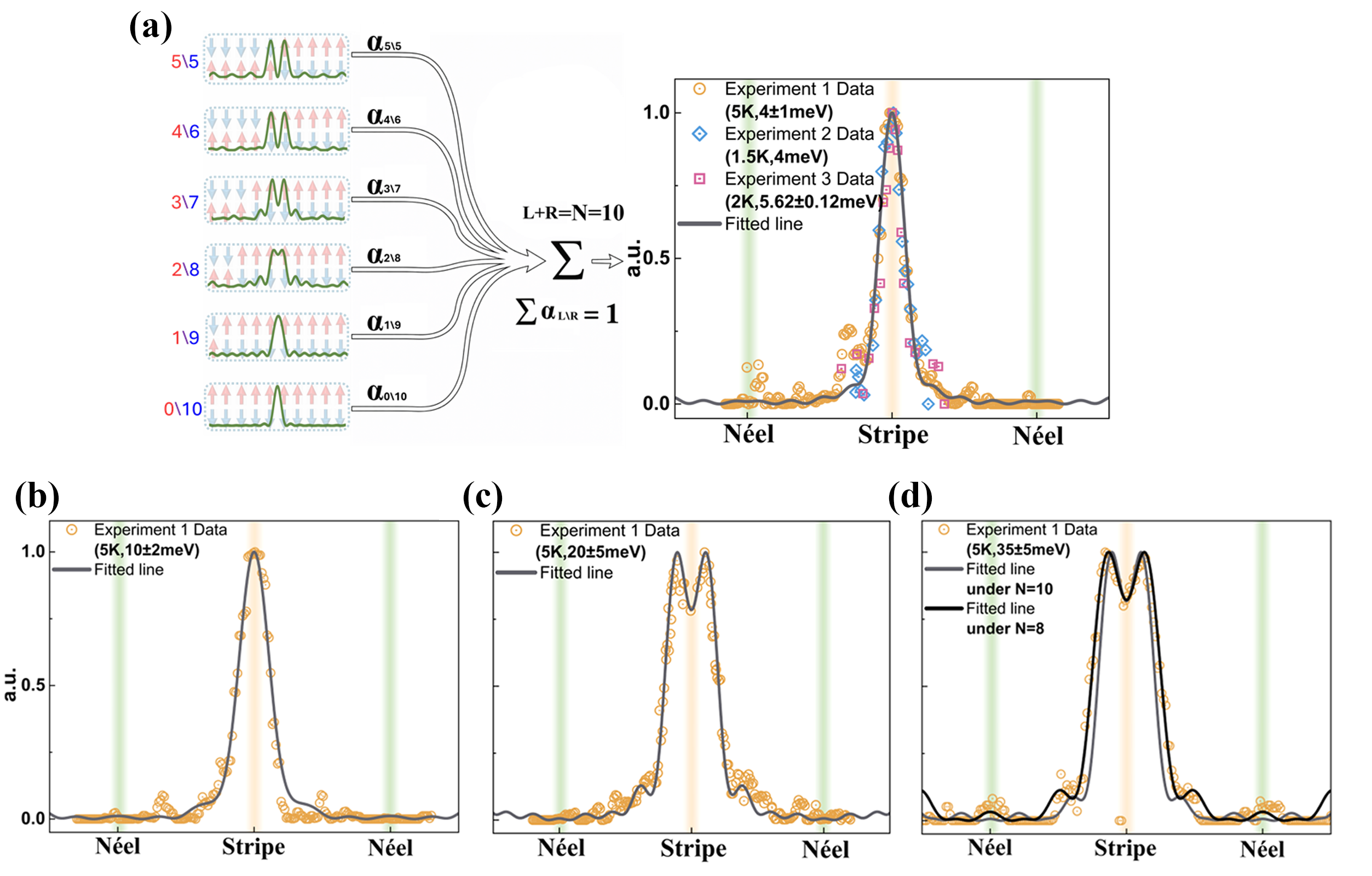}
    \caption{\textbf{Weighted reconstruction of INS line cuts.}
(a) $N=10$ weighted superposition of phase-slip configurations, $S_{\mathrm{fit}}(q)=\sum_{a\backslash b}\alpha_{a\backslash b}S_{a\backslash b}(q)$, compared with Experiment 1 \cite{ref8}, Experiment 2 \cite{ref6}, and Experiment 3 \cite{ref9}. The schematic labels the two segment lengths as $L$ and $R$, equivalent to $a$ and $b$ in the text. (b--d) Reconstructions of Experiment 1 at $E=10\pm2$, $20\pm5$, and $35\pm5$ meV; panel (d) compares $N=10$ and $N=8$.
}
    \label{fig:fig4}
\end{figure*}
The force field also carries a clear magnetic signature. Atoms at nearest-neighbor sites across the slip-wall interface experience a pronounced repulsive force, whereas next-nearest-neighbor sites feel an attractive force. This sign reversal shows that the slip wall is not a passive boundary inserted into an otherwise rigid stripe state. It is an active magnetoelastic object that reorganizes the local exchange environment while cutting the long-range stripe phase coherence. Therefore, an extended elastic response can coexist with a finite magnetic correlation length.

The slip-wall energetics reveal the second central ingredient: the wall selects its own finite scale [see Fig.~\ref{fig:fig3}(b)]. In N=10 and N=8 cases, slip walls with width 3 are all energetically favored [see Fig.~\ref{fig:fig2}(b)], whereas in the very short-period TRI(\(3\backslash3\)) state the same wall width carries a significant penalty. From configurations such as \(3\backslash3\), \(3\backslash5\), \(3\backslash7\), and \(3\backslash9\), we extract an effective formation energy for a single slip wall and find a minimum between \(3\backslash5\) and \(3\backslash9\). This gives an intrinsic magnetic correlation scale of about ten Fe spin sites for a width-3 slip wall. The energy lowering is not a trivial averaging effect from long stripe segments; it is a finite-coherence scale selected by the phase-slip energy landscape itself. This characteristic length directly controls the momentum linewidth of the INS scattering peaks.

The combined energy landscape [see Figs.~\ref{fig:fig2}(b) and \ref{fig:fig3}(b)] therefore selects a finite set of low-energy phase-slip configurations and suppresses the recovery of long-range magnetic coherence. The barrier near \(3\backslash9\) prevents a simple drift toward arbitrarily long stripe segments. This behavior is analogous to a nonlinear elastic response: near the optimal correlation length, the slip wall experiences a restoring constraint; beyond that scale, slip walls enter a more weakly coupled regime, while their formation energy remains higher than near the optimum \cite{ref28,ref29}. The resulting magnetic state is finite-range by construction. A successful reconstruction of the INS linewidth from this finite phase-slip set would then directly connect the real-space defect landscape to the observed momentum-space spectrum \cite{ref8,ref9}.

The final test is direct confrontation with INS line cuts. Existing itinerant/RPA approaches reproduce selected features of the FeSe spectrum but do not provide a unique real-space origin for the N\'eel-side response \cite{ref20,ref21}; Heisenberg-model-based descriptions leave key linewidth and spectral-weight features unresolved \cite{ref6,ref21,ref22}. Rather than attempting to solve the full dynamical structure factor $S(\mathbf q,\omega)$ for a fluctuating defect ensemble, we compare with the experimentally defined finite-energy-window response. For each local phase-slip cluster, we Fourier transform the real-space magnetic moments to obtain $S_{a\backslash b}(\mathbf q)$, and then form a nonnegative weighted superposition, $S_{\mathrm{fit}}(q)=\sum_{a\backslash b}\alpha_{a\backslash b}S_{a\backslash b}(q)$, as detailed in the Supplemental Material.

This construction is deliberately finite range. By partitioning the magnetic texture at the correlation scale $N$, it removes the artificial long-range coherence imposed by periodic boundary conditions and lets the linewidth emerge from the real-space size of the phase-slip clusters. Individual phase-slip configurations already generate a pronounced N\'eel-side response [see Fig.~\ref{fig:fig1}], whereas the stripe response is controlled by the density and size of the slip walls. Consistent with the energy hierarchy [see Fig.~\ref{fig:fig2}(b)], multiple phase-slip configurations remain accessible and must contribute statistically.

The weighted reconstruction reproduces the finite-energy-window INS line shapes in peak position, linewidth, splitting, and relative spectral-weight redistribution [see Fig.~\ref{fig:fig4}]. The N\'eel-side signal appears as the ensemble average of short-range N\'eel-like fragments generated by phase-slip fluctuations in a stripe background over the experimental timescale. The stripe channel, in turn, records the density and size of the slip walls through linewidth broadening and spectral-weight redistribution.

The three low-energy datasets compared [see Fig.~\ref{fig:fig4}(a)] were digitized from three different experimental papers. Experiment 1 was measured at $T=5$~K and $E=4 \pm 1$~meV \cite{ref8}; Experiment 2 at $T=1.5$~K and $E=4$~meV \cite{ref6}; and Experiment 3 at $T=2$~K and $E=5.625 \pm 0.125$~meV \cite{ref9}. The same $N=10$ finite-coherence set captures all three line cuts, including the narrower low-temperature profile of Experiment 2. This provides a direct experimental signature of ROSS: the spectrum remains organized by the same finite-coherence sector even when the detailed statistical weights vary between measurements.

The reconstruction is not only confined to the lowest-energy window. At \(E=10\pm2~\mathrm{meV}\) and \(E=20\pm5~\mathrm{meV}\), it captures the dominant stripe-centered response and the broadened line shape over the full momentum cut [see Figs.~\ref{fig:fig4}(b) and \ref{fig:fig4}(c)]. At higher energy, however, the full characteristic spectral range at $E=35\pm5$~meV is broader than the $N=10$ reconstruction [see Fig.~\ref{fig:fig4}(d)]. Combined with the slip-wall energetics [see Fig.~\ref{fig:fig3}(b)], the high-energy line shape is better represented by an effective shorter correlation length, $N=8$, whereas the barrier near \(3\backslash9\) suppresses substantial participation from $N=12$. Thus the ROSS texture ensemble is not static: its finite magnetic coherence is renormalized with excitation energy. FeSe is therefore governed by a nearly degenerate and dynamical phase-slip landscape, rather than by a single static ground state.

\section*{Conclusions}
{Taken together, our DFT calculations and spectrally weighted $S(\mathbf q)$ reconstruction establish FeSe as a phase-slip residual-order spin state (ROSS). Its magnetic spectrum is generated neither by a single ordered phase nor by random disorder, but by low-energy $\pi$ phase-slip textures embedded in a stripe background. Nonlocal magnetoelastic relaxation reduces the slip-wall cost, fixes a finite magnetic coherence scale, and converts local slip-wall structure into the observed principal stripe response, finite linewidth, peak splitting, and weak but reproducible N\'eel-side spectral weight.

More broadly, ROSS provides an organizing principle for frustrated quantum magnets in which local symmetry-broken correlations remain sharply defined while global phase coherence is destroyed by low-energy defects. In this sense, FeSe is not simply a magnet without order, but a prototype system realizing ROSS, with superconductivity developing on top of finite-coherence stripe magnetism.}

\section*{Data Availability}
The data that support the findings of this study are available from the corresponding author upon reasonable request.

\section*{Acknowledgments}
This work was supported by the
National Natural Science Foundation of China under Grant Nos. 12574254, 11974181,
11204131, and 12174059, and by the High Performance Computing Platform of
Nanjing University of Aeronautics and Astronautics. Some of the calculations were
performed at the High Performance Computational Center (HPCC) of the Department
of Physics at Fudan University.

\end{document}